\documentclass{llncs}
\usepackage{amsmath}
\usepackage{amsfonts}
\usepackage{amssymb}

\usepackage{graphicx}
\usepackage{todonotes}
\usepackage{enumitem}

\usepackage{theorem} 
\usepackage[driverfallback=hypertex]{hyperref} 

\newtheorem{lem}{Lemma}
\providecommand{\abs}[1]{\lvert #1 \rvert} 

\DeclareMathOperator{\var}{var}

\DeclareMathOperator{\aut}{Auk} 

\newcommand{\nc}{\newcommand}
\newcommand{\ol}{\overline}

\newcommand{\vp}{\varphi}

\newcommand{\ra}{\rightarrow}

\newcommand{\lra}{\leftrightarrow}

\newcommand{\sse}{\subseteq}

\newcommand{\fa}{\forall}
\newcommand{\ex}{\exists}

\newcommand{\mc}{\mathcal}

\newcommand{\set}[1]{\{ #1 \}}

\DeclareMathOperator{\nv}{N} 
\DeclareMathOperator{\na}{\nv_a} 
\DeclareMathOperator{\Aaut}{A}
\DeclareMathOperator{\Eaut}{E}

\nc{\bm}{\boldmath}
\nc{\bmm}[1]{\mbox{\bm$\DST #1$}}

\theoremstyle{definition}

\let\llncssubparagraph\subparagraph
\let\subparagraph\paragraph
\usepackage[compact]{titlesec}
\let\subparagraph\llncssubparagraph
\titlespacing\subsection{0pt}{12pt plus 4pt minus 2pt}{0pt plus 2pt minus 2pt}

\begin{document}

\title{Introducing Autarkies for DQCNF}

\author{Oliver Kullmann\inst{1}\thanks{Supported by EPSRC grant EP/S015523/1.} \and
Ankit Shukla\inst{2}}
\institute{Computer Science Department\\
		Swansea University, UK\\
\email{O.Kullmann@Swansea.ac.uk}\\
 \and
{Institute for
Formal Models
and Verification\\
Johannes Kepler University, Austria}\\
\email{ankit.shukla@jku.at}}

\maketitle

\begin{abstract}
Autarkies for SAT can be used for theoretical studies, preprocessing and inprocessing. 
They generalise satisfying assignments by allowing to leave some clauses ``untouched" (no variable assigned). 
We introduce the natural generalisation to DQCNF (dependency-quantified boolean CNF), with the perspective of SAT translations for special cases.
Finding an autarky for DQCNF is as hard as finding a satisfying assignment. 
Fortunately there are (many) natural autarky-systems, which allow restricting the range of autarkies to a more feasible domain, while still maintaining the good general properties of arbitrary autarkies. 
We discuss what seems the most fundamental autarky systems, and how the related reductions can be found by SAT solvers.

\end{abstract}

\section{Introduction}
\label{sec:intro}

The theory of autarkies for SAT (in CNF)
(see \cite{Kullmann2007HandbuchMU} for an overview) is used in the study of unsatisfiability and pre/inprocessing (\cite{LiffitonSakallah2008Trimming} is an early paper).
In this work we introduce basic autarky theory for DQBF (see \cite{KBB09HBSAT} for QBF).
Indeed we concentrate on DQCNF. These formulas allow existential quantifiers with specified dependencies on universal variables, where the matrix is a CNF.
An autarky for a CNF $F$ is a partial assignment which satisfies every clause of $F$ containing an assigned variable.
Clauses satisfied by some autarky can be removed
satisfiability-equivalently.
This is generalised to DQCNF by considering partial assignments to existential variables and allowing boolean functions as values (fulfilling the dependencies), where the clauses with assigned variables need now to become tautologies.
So we start right with the boolean functions in this semantical perspective, and do not extract them as in  \cite{BalabanovChiangJiang2014DQBF,WWSB2016DQBF}.
Restricted forms
of autarkies yield “autarky systems (see \cite[Section 11.11]{Kullmann2007HandbuchMU}).
We present three autarky systems for DQCNF, namely, $\Aaut_0$, $\Aaut_1$ and $\Eaut_1$ (using ``A'' to denote universal variables, and ``E'' for existential variables).
The basic autarky-systems $\Aaut_0$, $\Aaut_1$ allow the boolean functions to \emph{essentially} depend on $0$ resp.\ $1$ universal variable, while $\Eaut_1$ only uses one existential variable (for a single autarky).
Consider the example
\begin{eqnarray*}
  F & := & \fa x_1,x_2,x_3 \ex y_1(x_1,x_2) \ex y_2(x_2,x_3) \ex y_3(x_1): F_0\\
  F_0 & := & (y_1 \vee x_1) \wedge (\overline{y_1} \vee x_2) \wedge (\overline{y_2} \vee \overline{x_2} \vee x_3) \wedge (y_3 \vee \overline{x_1} \vee x_2) \wedge (\overline{y_3} \vee x_1).
\end{eqnarray*}
Since $\overline{y_2}$ is pure, we have the $\Aaut_0$-autarky $y_2 \rightarrow 0$ (removing the third clause).
Furthermore we have the $\Aaut_1$-autarky $y_3 \rightarrow x_1$, removing the fourth and fifth clauses.
Both these autarkies are also $\Eaut_1$-autarkies.
We obtain the reduction result $\fa x_1,x_2,x_3 \ex y_1(x_1,x_2) : (y_1 \vee x_1) \wedge (\overline{y_1} \vee x_2)$, which is equivalent to the QBF $\fa x_1,x_2 \ex y_1 : (y_1 \vee x_1) \wedge (\overline{y_1} \vee x_2)$, which doesn't allow any further autarky. Such a final result of autarky reduction is called the (unique) \textbf{lean kernel} of the original formula.

\section{Confluence of autarky reduction}
\label{sec:autsys}

A DQCNF is called \textbf{lean} if it has no non-trivial autarkies.
The union of two lean DQCNF with compatible variables and dependencies is again lean, and thus every DQCNF has a largest lean sub-DQCNF, the \textbf{lean kernel}.
We denote the lean kernel by $\bm{\na(F)}$ (``N'' like ``normal form'').
Alternatively one can arrive at the lean kernel via \textbf{autarky reduction}.
For an autarky $\vp$ of a DQCNF we denote by $\vp * F$ the DQCNF with the clauses removed which are satisfied by $\vp$.
\begin{lem}\label{lem:autsateq}
  $\vp * F$ is satisfiability-equivalent to $F$ for an autarky $\vp$ of $F$.
\end{lem}
For two autarkies $\vp, \psi$ of $F$ one can consider the composition $\vp \circ \psi$, which on the variables of $\psi$ acts like $\psi$, and otherwise like $\vp$:
\begin{lem}\label{lem:compaut}
  The composition of two autarkies is again an autarky.
\end{lem}
Now the lean kernel is obtained by repeatedly applying autarky-reduction on $F$ as long as possible:
\begin{lem}\label{lem:decomp}
  Consider a DQCNF $F$. The largest lean sub-DQCNF $\na(F)$ is also obtained by applying autarky-reduction to $F$ as long as possible (in any order).
\end{lem}

Due to the high complexity of general autarky-finding for DQCNF, it is vital to allow restricted notions of autarkies (with lower complexity).
For that purpose we generalise the notion of ``autarky systems'' from \cite[Section 11.11]{Kullmann2007HandbuchMU}.
For a DQCNF $F$ we write $\aut(F)$ for the set of all autarkies.
An \textbf{autarky system} $\mc{A}$ allows to consider subsets $\mc{A}(F) \sse \aut(F)$.
There are two basic conditions which make $\mc{A}$ an autarky system.
First for DQCNFs $F$ and $\vp, \psi \in \mc{A}(F)$ it must always hold $\vp \circ \psi \in \mc{A}(F)$ (closure under composition).
And second for DQCNFs $F \sse F'$ it must always hold $\mc{A}(F') \sse \mc{A}(F)$ (removal of clauses does not remove $\mc{A}$-autarkies).
\textbf{$\mc{A}$-satisfiability} means satisfiability by a series of $\mc{A}$-autarkies, while \textbf{$\mc{A}$-leanness} means there there are no nontrivial $\mc{A}$-autarkies.
To have Lemma \ref{lem:decomp} (and further natural properties) the following four conditions on $\mc{A}$ are fundamental:
\begin{description}
\item[standardised] variables not actually occurring are irrelevant.
\item[$\bot$-invariant] universal clauses (only having universal variables) are irrelevant.
\item[invariant under variable elimination] removing (existential) variables from the clauses does not affect autarkies of $\mc{A}$ which do not use these variables.
\item[invariant under renaming] renaming variables (existential or universal) is respected by the autarky system.
\end{description}

These four conditions are always expected to hold.
Additionally we call $\mc{A}$ \textbf{iterative}, if we have that if we apply any autarky of $\mc{A}$, and take another autarky of $\mc{A}$ for the reduction result, then their composition also belongs to $\mc{A}$ (for the original DQCNF).
For iterative systems every result from autarky reductions can also be obtained by applying a \emph{single} autarky from the system.
Autarky systems fulfilling all five conditions are called \textbf{normal}.

\section{A- and E-systems}
\label{sec:AES}

Consider a DQCNF $F$ and $k \ge 0$:
\begin{itemize}
\item An \textbf{$\Aaut_k$-autarky} for $F$ is an autarky such that all boolean functions assigned depend essentially on at most $k$ variables.
\item An \textbf{$\Eaut_k$-autarky} is an autarky assigns at most $k$ (existential) variables.
\end{itemize}
$\Aaut_0$ is just CNF-autarky, and we concentrate on $\Aaut_1$.
Generalising \cite{HKBSubramaniZhao2003BooleanModels} one sees that all satisfiable DQ2CNF are satisfiable by an $\Aaut_1$-assignment.
The composition $\mc{E}_1 + \mc{A}_1$ one might consider as the basic ``clean-up autarky system''.

Since the existence of an ordinary propositional autarky for a clause-set $F$ is NP-complete (see \cite{Ku00f}), which is covered by $\Aaut_0$-autarkies, deciding whether a DQCNF has a non-trivial $\Aaut_k$-autarky is NP-complete for every $k \ge 0$ (boolean functions can be just represented by truth-tables here).
Deciding the existence and finding some short $\Eaut_1$-autarky can be done in polynomial time (discussed below).
The $\Aaut_k$-systems are all normal, while the $\Eaut_k$ fulfil the basic conditions, but are not iterative for $k \ge 1$.

\subsection{$\Eaut_1$-autarkies}
\label{sec:e1aut}

Consider any DQCNF $F$ and a set $V$ of existential variables.
Let the DQCNF $F[V]$ be obtained by restriction to $V$, removing all clauses not containing some variable from $V$, and removing from the remaining clauses all existential literals with variables not from $V$.
The existence of any autarky $\vp$ with given var-set $\var(\vp) = V$ is equivalent to the DQCNF $F[V]$ being satisfiable, where the satisfying assignments correspond directly to the autarkies.

So searching for an $\Eaut_1$-autarky of $F$ can be done by solving for each $v$ the one-existential-variable DQCNF $F[\set{v}]$.
This can be done easily, since in general from any clause $C$ we can remove universal literals with variables not in any domain of some existential variable of $C$.
So we have a QCNF with prefix $\fa\ex$ and with exactly one existential variable to solve.
It is easy to see that the solution-set for $v$ is an interval $A \ra v \ra B$, which can be read off from the positive resp.\ the negative occurrences of $v$ (as DNF resp.\ CNF), and where choosing for $v$ either $A$ or $B$ yields easy solutions.
We note here that for $\Eaut_2$ we do not get a QCNF with two existential variables, but only such a DQCNF.

\subsection{$\Aaut_1$-autarkies}
\label{sec:a1aut}

The main strategy to solve the constraint-satisfaction problem of finding some non-trivial $\Aaut_1$-autarky for a given $F$ is: 1. to explicitly list the possible boolean functions as values of the existential variables, 2. to \emph{compile} for each clause $C \in F$ the minimal possibilities for $C$ to become a tautology, and 3. to make it explicitly part of the SAT-translation that at least one of these \emph{minimal} possibilities for $C$ is fulfilled, if the selector-variable of $C$ is true (non-triviality of the autarky means that at least one selector-variable is true).
For an existential variable $v$ with dependency-set $D(v)$ we thus have $2 \abs{D(v)} + 2$ possible values to consider, the two constant boolean functions plus for each universal variable the two associated literals.
And there are exactly three types of minimal possibilities for $C$ to become a tautology:
\begin{enumerate}
\item some existential $y \in C$ is set to $1$;
\item for some existential $y \in C$ there is a universal $x \in C$ with $\var(x) \in D(\var(y))$, and $y$ is set to $\ol x$;
\item for some existential $y, y' \in C$, $y \ne y'$, there is a universal $x \in D(\var(y))$, and $y$ is set to $x$ and $y'$ to $\ol x$.
\end{enumerate}
The translation into a SAT-problem is now in principle not difficult, but effort is needed for the representation of AMO-constraints (at-most-one), and for encodings of nonboolean values (direct or logarithmic encodings).

Obviously the strength of $\Eaut_1$ and $\Aaut_1$ in general is incomparable. On the formula in the Introduction indeed the $\Eaut_1$-lean kernel is the same as the $\Aaut_1$-lean kernel, namely the (absolute) lean kernel.
But for example for the following formula they produce different normalforms:
\begin{eqnarray*}
	F & := & \fa x_1 \ex y_1() \ex y_2(x_1): F_0\\
	F_0 & := & (y_1 \vee \overline{y_2} \vee x_1) \wedge (y_2 \vee x_1) \wedge (y_1 \vee y_2 \vee \overline{x_1}) \wedge (\overline{y_1} \vee \overline{y_2} \vee \overline{x_1}).
\end{eqnarray*}
First consider $\Eaut_1$. Variable $y_1$ does not yield something, since not pure. While variable $y_2$ alone can only be set to true or $\overline{x_1}$, due to the second clause, while due to the first and third clause this does not yield an autarky. So $F$ is $\Eaut_1$-lean.
But this formula is $\Aaut_1$-satisfiable with assignment $y_1 \mapsto 1$ and $y_2 \mapsto \ol{x_1}$. For the opposite consider $\fa\,x_1,x_2 \ex y : y \lra x_1 \vee x_2$, which is $\Eaut_1$-satisfiable, but $\Aaut_1$-lean.

With improved SAT-encodings and running standard SAT-solvers, we where able to compute the normalforms quite easily for all over 334 instances in the DQBF track of QBFEVAL'18 (\cite{Qbfeval18}).

\section{Conclusion}
\label{sec:conclusion}

We consider $\Eaut_1 + \Aaut_1$ only as a first example and appetiser --- all the SAT-theory on autarkies can be combined with many interesting classes of boolean functions, to yield interesting autarky systems.
The results on the normalforms by the basic autarky systems $\Aaut_0, \Aaut_1, \Eaut_1$ and their combinations, on all known QCNF and DQCNF instances (as in QBFLIB), will be made available online.
After establishing this precise basis, the use of autarkies in pre- and inprocessing is the main question.
Theoretical studies of lean DQCNF (see \cite{BueningZhao2008QBFFixDef} for using $\Aaut_0$ for QBF) should also be very interesting.

\bibliographystyle{splncs04}
\bibliography{2019_QBF_WORKSHOP}

\end{document}